**Title: Cross-Attention Multimodal Learning for Predicting Response to Neoadjuvant Imatinib in Gastrointestinal Stromal Tumors: A Multicenter Retrospective Study**

**Authors:** Fariba Tohidinezhad[1,3,*], Douwe J. Spaanderman[1], Natalia Oviedo Acosta[1,3], Kaouther Mouheb[1], Karthik Prathaban[1,3], David F. Hanff[1], Dirk J. Grünhagen[2], Cornelis Verhoef[2], Joris M. van Sabben[4], Evelyne Roets[4], Jette J. Slettenhaar[1], Hans Gelderblom[5], Ingrid M.E. Desar[6], Anna K.L. Reyners[7], Neeltje Steeghs[8], Stefan Klein[1,9], Martijn P.A. Starmans[1,3,9]

**Affiliations:**

1. Department of Radiology and Nuclear Medicine, Erasmus MC Cancer Institute, University Medical Center Rotterdam, Rotterdam, the Netherlands.
2. Department of Surgical Oncology, Erasmus MC Cancer Institute, University Medical Center Rotterdam, Rotterdam, the Netherlands.
3. Department of Pathology, Erasmus MC Cancer Institute, University Medical Center Rotterdam, Rotterdam, the Netherlands.
4. Department of Medical Oncology, The Netherlands Cancer Institute, Amsterdam, The Netherlands.
5. Department of Medical Oncology, Leiden University Medical Center, Leiden, the Netherlands.
6. Department of Medical Oncology, Radboud University Medical Centre, Nijmegen, the Netherlands.
7. Department of Medical Oncology, University Medical Center Groningen, University of Groningen, Groningen, the Netherlands.
8. Department of Medical Oncology, Netherlands Cancer Institute, Antoni Van Leeuwenhoek, Amsterdam, the Netherlands.
9. Shared last author.

**Corresponding author:** Fariba Tohidinezhad, Department of Radiology and Nuclear Medicine, Erasmus MC Cancer Institute, University Medical Center Rotterdam, Dr. Molewaterplein 40, 3015 GD Rotterdam, The Netherlands. Email: f.tohidinezhad@erasmusmc.nl.

**Funding:** This work is part of the AIID project, funded by the Dutch Research Council (NWO) under the AiNed Fellowship Grants program (project number NGF.1607.22.025). Additional support was provided by an unrestricted grant from the Stichting Hanarth Fonds.

**Acknowledgment:** This work used the Dutch national e-infrastructure with the support of the SURF Cooperative using grant no. EINF-14257, financed by the Dutch Research Council (NWO). We gratefully acknowledge the fundraising efforts of patients, families, and supporters from the GIST community, including a cycling initiative led by Rob Scheurink and collaborators, which contributed to the development of the imaging dataset.

**Abstract**

**Background:** Response to neoadjuvant imatinib in gastrointestinal stromal tumors (GISTs) is highly variable and cannot be reliably predicted using current clinical or molecular markers. This study developed and evaluated an explainable multimodal deep learning framework integrating computed tomography (CT) imaging and clinical variables to predict treatment response.

**Methods:** Patients from four tertiary centers were retrospectively included between 2000-2023 in independent pretraining (n=935) and prediction (n=213) cohorts. A cross-attention framework integrating clinical variables and tumor-centered CT imaging was developed to predict response to neoadjuvant imatinib. Two training strategies were evaluated: (1) self-supervised pretraining with low-rank adaptation and (2) training from scratch. Hyperparameters were optimized using SMAC3. Performance was assessed through internal cross-validation and external testing. Ablation analyses and attention-based explanations were used to quantify modality contributions.

**Results:** Among 213 patients (54.5% responders), responders had larger tumors (112 vs. 89 mm, P=0.026), higher mitotic index (3 vs. 0, P<0.001), and more frequent KIT mutations (69.0% vs. 56.7%, P=0.019). Cross-attention models achieved the highest internal performance (AUC up to 0.99) but lower external performance (AUC 0.60-0.63). Clinical-only performance was moderate (AUC 0.66), whereas imaging-only models showed limited generalizability (AUC 0.56-0.66). Explainability analyses identified significant differences in feature importance between responders and non-responders, including CD117, BRAF, PDGFRA, age, sex, disease status, and comorbidities (FDR-adjusted P<=0.036).

**Conclusion:** The cross-attention framework shows potential for improving imatinib response prediction in GIST while providing interpretable insights into multimodal determinants of treatment response.

## Background

Gastrointestinal stromal tumors (GISTs) are the most common mesenchymal neoplasms of the gastrointestinal tract, with an estimated incidence of 10–15 cases per million per year[1]. They arise predominantly in the stomach (60–70%) and small intestine (20–30%) and originate from the interstitial cells of Cajal or related precursor cells[1, 2]. At the molecular level, the majority of GISTs are driven by activating mutations in the receptor tyrosine kinases KIT or platelet-derived growth factor receptor alpha (PDGFRA), resulting in constitutive signaling and uncontrolled cellular proliferation[3]. Surgical resection remains the mainstay of treatment for localized disease[4]. The introduction of tyrosine kinase inhibitors (TKIs), particularly imatinib, has markedly improved outcomes in advanced and metastatic settings[5]. In addition, neoadjuvant imatinib has been increasingly employed in patients with locally advanced or borderline resectable tumors to reduce tumor size and facilitate less morbid, organ-preserving surgery[4, 6].

Despite the clinical success of imatinib, response to therapy remains highly heterogeneous across patients, with reported objective response rates of approximately 50–70% in advanced disease[7]. While certain molecular subtypes, such as KIT exon 11 mutations, are generally associated with favorable responses, others, including KIT exon 9 mutations or PDGFRA D842V mutations, demonstrate reduced sensitivity or primary resistance[8]. However, mutational status alone does not fully explain the observed variability in treatment outcomes, and patients with similar molecular profiles may exhibit markedly different responses. Currently, there is no reliable, non-invasive tool to predict individual response to neoadjuvant imatinib prior to treatment initiation. Consequently, some patients may be exposed to ineffective therapy, leading to delays in definitive surgical management, unnecessary toxicity, and potential disease progression[4, 9].

Multimodal learning (MML) has emerged as a compelling framework for integrating complementary sources of patient information[10, 11]. Clinical decision-making inherently draws on diverse data modalities, including imaging and clinical characteristics, each capturing distinct facets of tumor biology and patient heterogeneity, reflecting the multidisciplinary nature of oncologic practice where such information is jointly considered to guide treatment strategies. Nevertheless, many existing predictive approaches remain limited to a single modality or rely on simplistic fusion techniques, such as feature concatenation, which fail to adequately capture the complex interdependencies between data sources. Recent advances in deep learning have enabled more expressive multimodal architectures capable of jointly modeling heterogeneous inputs. Cross-attention mechanisms, provide a powerful means of capturing interactions between modalities by allowing representations from one domain to selectively attend to the most relevant information in another, while inherently offering interpretability by elucidating modality-specific contributions[12–14].

This study aimed to: (1) develop and externally evaluate a cross-attention–based multimodal deep learning framework for predicting response to neoadjuvant imatinib in patients with locally advanced GIST, integrating three-dimensional computed tomography (CT) imaging with clinical variables; (2) systematically compare training paradigms, including self-supervised pretraining with transfer learning and training from scratch; (3) assess the relative contribution of imaging and clinical modalities through ablation analyses; and (4) characterize model explainability by examining clinical feature importance and imaging attention patterns.

## Methods

*Study Population and Data Collection*

Patients were retrospectively included in two independent multicenter cohorts: (a) a pretraining cohort for representation learning; and (b) a prediction cohort for response to neoadjuvant imatinib (Figure 1). Data were collected from four tertiary referral centers in the Netherlands: the Netherlands Cancer Institute–Antoni van Leeuwenhoek Hospital (AVL), Erasmus MC Cancer Institute (EMC), Radboud University Medical Center (RUMC), and Leiden University Medical Center (LUMC). The pretraining cohort comprised patients diagnosed with GIST between 2000 and 2023 (AVL, n=364; EMC, n=231; RUMC, n=162; LUMC, n=178), whereas the prediction cohort included patients treated with neoadjuvant imatinib between 2008 and 2023 (AVL, n=118; EMC, n=48; RUMC, n=26; LUMC, n=21). To ensure patient-level independence, individuals in the prediction cohort were excluded from the pretraining cohort.

Eligible patients in the pretraining cohort had a confirmed GIST diagnosis and available baseline CT imaging. For the prediction cohort, inclusion required neoadjuvant imatinib treatment, a contrast-enhanced CT scan within four months prior to treatment initiation, and documented response evaluation. Rectal GISTs were excluded from the prediction cohort due to the preferential use of magnetic resonance imaging for local staging and the suboptimal delineation of rectal tumors on CT. Imaging was performed according to standard-of-care protocols across centers, in line with European Society for Medical Oncology and German GIST Imaging Working Group recommendations[4, 15], generally including abdominal CT in the portal venous phase with thin-slice reconstruction and multiplanar reformations; thoracic CT was performed when clinically indicated.

Treatment response was assessed according to RECIST version 1.1 and dichotomized into responders (complete or partial response) and non-responders (stable or progressive disease)[16]. Clinical variables included demographic characteristics (age, sex), tumor-related features (tumor status at diagnosis, primary site, longest axial diameter, histology, and mitotic index, immunohistochemical markers (CD117 and DOG1), molecular alterations (KIT, PDGFRA, and BRAF mutation status), and comorbidities (diabetes mellitus, hypertension, hypercholesterolemia, and history of other malignancies). The study was conducted in accordance with the Declaration of Helsinki and approved by the institutional review boards of all participating centers, with waiver of informed consent due to its retrospective nature.

*Tumor Segmentation*

For the prediction cohort, primary tumors were segmented on baseline CT scans to enable extraction of tumor-specific imaging representations. Segmentation was performed using nnInteractiveNet, a promptable deep learning–based framework for 3D medical image segmentation, implemented via the 3D Slicer extension (SlicerNNInteractive)[17, 18]. All segmentations were conducted by a single annotator (F.T.). To assess inter-observer variability, a second independent annotator (D.S.) segmented a randomly selected 20% subset using the same protocol. Agreement between annotators was quantified using the Dice similarity coefficient (DSC). To further characterize boundary agreement, the 95th-percentile Hausdorff distance and the average symmetric surface distance were computed. In addition, volumetric agreement was assessed using volume similarity.

*Data Processing*

CT images and corresponding tumor segmentations were resampled to a common voxel spacing, intensity-normalized, and padded to fixed tumor-centered volumes. During training, random in-plane rotations were applied for augmentation. Volumes were subsequently partitioned into non-overlapping 3D patches to form imaging tokens, with padding-aware masking used to exclude non-informative regions. Clinical variables were encoded as a fixed-length token sequence through discretization of continuous features, categorical encoding, and explicit handling of missing values, with additional stochastic masking applied during training.

*Multimodal Cross-Attention Architecture*

A bidirectional cross-attention Transformer was used to fuse clinical and imaging representations at the token level[19]. Modality-specific self-attention layers captured intra-modal dependencies, followed by reciprocal cross-attention blocks enabling mutual conditioning between clinical and imaging tokens. The fused representations were aggregated via attention pooling to derive a patient-level embedding, which was used for binary response prediction. The model was implemented using the FuseMedML framework[20].

*Training Strategies*

The training framework, encompassing self-supervised representation learning and downstream model optimization, is illustrated in Figure 2.

*Self-Supervised Pretraining of the Imaging Encoder:* Masked autoencoding (MAE) was used for self-supervised pretraining with an asymmetric encoder–decoder architecture[21]. Anatomically localized volumetric crops (64 × 256 × 256 voxels) were extracted by restricting inputs to the abdominal region using automatically generated body and liver masks (TotalSegmentator, v2.12.0)[22], with the superior-most liver slice defining the upper landmark. Inputs were partitioned into non-overlapping patches, and 75% of valid tokens were randomly masked. The encoder processed only visible tokens, while a lightweight decoder reconstructed the full volume from latent representations and mask tokens. Optimization minimized mean squared reconstruction error over masked tokens.

*Fine-Tuning:* For transfer learning, the imaging encoder was initialized with the self-supervised pretrained weights and adapted using low-rank adaptation (LoRA)[23, 24]. The pretrained encoder weights were frozen, and trainable low-rank adapters were injected into the query and value projection matrices of the imaging self-attention layers. The remaining trainable components were optimized using AdamW with binary cross-entropy loss.

*Training from Scratch:* In the training-from-scratch setting, the imaging encoder was initialized with random weights and trained directly from raw tumor-centered inputs under supervision. In contrast to the transfer learning approach, the imaging encoder parameters were optimized jointly with the remaining model components.

*Hyperparameter Optimization*

Hyperparameters were optimized using SMAC3 (Sequential Model-based Algorithm Configuration version 3)[25] under a five-day wall-time budget on a single NVIDIA A100 GPU. SMAC was initialized with a Sobol design comprising 10 initial configurations, followed by Bayesian optimization using a 10-tree random forest surrogate model and expected improvement (EI) as the acquisition function, with a random-design probability of 0.2 to maintain exploration. Each configuration was evaluated using five-fold cross-validation on the development cohort (AVL, RUMC, and LUMC), stratified by center and response label, with the best validation epoch selected per fold. The optimization objective was defined as $Cost = 1 - \frac{1}{K}\sum_{k=1}^{K} AUC_k$, where K=5. The search space comprised optimization, multimodal architecture (clinical encoder, imaging encoder, and cross-attention modules), classifier head, and data-processing and augmentation hyperparameters, with additional strategy-specific parameters depending on the training paradigm. Full search spaces are provided in Supplementary Tables S1–S2 (fine-tuning) and Tables S3–S5 (training from scratch). Final predictions were obtained by ensembling the top five distinct configurations using mean-logit aggregation.

*Performance Assessment*

Internal validation was conducted using out-of-fold predictions on the development cohort. For each fold, predictions were generated by ensembling fold-matched models from the top five configurations, excluding models trained on that fold; logits were averaged to obtain final probabilities. External performance was assessed on the independent EMC cohort, held out from training and model selection. Discrimination was evaluated using the area under the receiver operating characteristic curve (AUC), sensitivity, specificity, positive predictive value (PPV), negative predictive value (NPV), and accuracy. Calibration was assessed using expected calibration error (ECE) with corresponding reliability diagrams[26].

*Ablation analysis*

Unimodal ablations were conducted to quantify modality-specific contributions. Clinical-only models encoded tokenized clinical variables using a Transformer, whereas imaging-only models processed tumor-centered 3D patch tokens via the imaging encoder, without cross-attention. Each ablation was optimized under an identical evaluation and optimization framework, with modality-specific adaptations, ensuring a controlled and directly comparable assessment of the incremental value of multimodal fusion.

*Explainability*

To characterize modality-specific contributions, post hoc attention-based analyses were performed at both clinical and imaging levels[27]. For clinical variables, token-level attention weights were aggregated across layers and heads and ranked to quantify relative feature importance. In addition, attention distributions were compared between response and non-response groups to assess differential feature importance. Statistical significance was evaluated using the Mann–Whitney U test, with p-values corrected for multiple comparisons using the Benjamini–Hochberg false discovery rate (FDR) method. For imaging, attention weights corresponding to tumor-centered patch tokens were projected back to the spatial domain and overlaid on corresponding CT slices to localize regions driving model predictions. This approach enables complementary interpretation of tabular and imaging modalities within a unified framework.

## Results

The prediction cohort comprised 213 patients, of whom 116 (54.5%) achieved an objective response to neoadjuvant imatinib and 97 (45.5%) were classified as non-responders (Table 1). Treatment response rates were comparable across the four participating centers, ranging from 52.6% to 58.7%. The median age at treatment initiation was 65 years [56–73], and 55.9% were male. Most tumors originated in the stomach (70.4%) and were locally advanced at diagnosis (87.8%).

Several clinicopathological variables differed between responders and non-responders. Responders more frequently presented with locally advanced disease (94.8% vs. 79.4%), whereas localized disease was more common among non-responders (16.5% vs. 4.3%; P=0.003). Responders also exhibited a higher tumor burden, with a larger longest axial diameter (112 [76–160] mm vs. 89 [60–140] mm, P=0.026) and a higher mitotic index (3 [0–9.2] vs. 0 [0–1], P<0.001). In addition, KIT mutations were more prevalent among responders (69.0% vs. 56.7%), while absence of KIT mutations was more frequent in non-responders (P=0.019). No significant differences were observed for age or sex, although responders tended to be younger and more frequently female (P=0.086 and P=0.081, respectively). Other variables, including tumor site, histology, immunohistochemical markers, and comorbidities, were not significantly different between groups.

The development and external test cohorts were broadly comparable (Supplementary Table S6), with no significant differences in response rates (43.6% vs. 52.1%, P=0.4), demographic characteristics, tumor location, or disease stage. Notable differences included a lower mitotic index in the external test cohort (0 [0–1.8] vs. 1 [0–6], P=0.043) and a lower prevalence of KIT mutations (52.1% vs. 66.7%, P=0.044), alongside a higher proportion of missing molecular data. A trend toward differences in histological subtype distribution was also observed (P=0.087).

Inter-observer agreement for tumor segmentation was assessed in 43 cases. Agreement between annotators yielded a median DSC of 0.96 [0.92–0.98]. Boundary-based metrics further supported strong concordance, with a median 95th-percentile Hausdorff distance of 3.84 mm [2.30–5.62] and an average symmetric surface distance of 0.65 mm [0.51–1.20]. Volume similarity showed a median of 0.987 [0.950–0.995], indicating consistent tumor delineation between observers.

Self-supervised learning was conducted on the multicenter pretraining cohort of 935 patients. The MAE exhibited stable convergence over 200 epochs, with reconstruction loss decreasing from 0.26 to 0.055 on the training set and from 0.23 to 0.054 on the validation set (Supplementary Figure S1). Training and validation losses remained closely aligned throughout optimization, with no observable divergence. Loss reduction was most pronounced during early epochs, followed by gradual plateauing, consistent with stable optimization dynamics.

The hyperparameter optimization objective varied across model paradigms, with at least 57 configurations evaluated per model (Supplementary Figure S1). The fine-tuned cross-attention model spanned 0.25–0.35, while the fine-tuned imaging-only model exhibited substantially higher values, ranging from 0.47 to 0.57. Models trained from scratch showed broader distributions, with cross-attention spanning 0.21–0.43 and imaging-only ranging from 0.31 to 0.57. The clinical-only model demonstrated a narrower range of 0.30–0.37 across configurations.

The top five configurations per model are presented in Supplementary Tables S7–S11. The cross-attention model trained from scratch reaching the highest mean validation AUCs (0.734–0.788), followed by the fine-tuned cross-attention model (0.737–0.750). The clinical-only model achieved intermediate performance (0.680–0.696), whereas imaging-only models showed lower AUCs, both for fine-tuning (0.510–0.525) and training from scratch (0.634–0.693).

Internal and external validation performance are summarized in Figure 3 and Supplementary Table S12. In internal validation, cross-attention models showed the highest discrimination. The from-scratch cross-attention model achieved an AUC of 0.99 (95% CI, 0.99–1.00), with high accuracy (0.93 [0.89–0.97]), sensitivity (1.00 [1.00–1.00]), and specificity (0.88 [0.81–0.95]), followed by the fine-tuned counterpart (AUC, 0.94 [0.91–0.97]; accuracy, 0.87 [0.81–0.92]). The clinical model yielded moderate performance (AUC, 0.87 [0.82–0.92]). Imaging-only models were inconsistent, with absent discrimination in the fine-tuned setting and near-perfect performance when trained from scratch. Calibration varied across models (Brier score, 0.01–0.25; ECE, 0.07–0.18).

In external validation, discrimination decreased across all models. The fine-tuned cross-attention model achieved an AUC of 0.63 (0.47–0.78) with balanced sensitivity (0.56 [0.36–0.76]) and specificity (0.57 [0.35–0.78]), while the from-scratch variant showed comparable AUC (0.60 [0.43–0.75]) with higher specificity (0.70 [0.52–0.87]) and lower sensitivity (0.52 [0.32–0.72]). The clinical model demonstrated similar discrimination (AUC, 0.66 [0.50–0.80]). Imaging-only models showed limited generalizability (AUC, 0.56–0.66), including a degenerate classification pattern in the fine-tuned setting. Calibration remained modest in external validation (Brier score, 0.24–0.30; ECE, 0.01–0.30).

Explainability analyses revealed the contributions from clinical and imaging modalities (Figure 4). At the clinical level, several features showed significant differences in attention-based importance between responders and non-responders after FDR correction (Figure 4A). Features with higher importance in responders included molecular markers, namely CD117 (mean importance 0.0635 vs. 0.0632; FDR-adjusted $P=0.010$), BRAF (0.0633 vs. 0.0600; $P=0.008$), and PDGFRA (0.0626 vs. 0.0605; $P=0.031$). In contrast, multiple clinical and demographic variables demonstrated higher importance in non-responders, including status at diagnosis (0.0642 vs. 0.0632; $P=0.008$), age at imatinib initiation (0.0636 vs. 0.0622; $P<0.001$), sex (0.0633 vs. 0.0624; $P=0.025$), and comorbidity-related features such as other malignancies (0.0632 vs. 0.0619; $P=0.003$), hypertension (0.0629 vs. 0.0620; $P=0.036$), hypercholesterolemia (0.0628 vs. 0.0617; $P=0.004$), and diabetes (0.0627 vs. 0.0617; $P=0.025$). At the imaging level, attention maps projected onto tumor-centered CT slices localized salient regions within the tumor volume that contributed to predictions. Representative external cases illustrated distinct spatial attention patterns, with focused regions driving predictions in both true positive and true negative cases (Figure 4B).

## Discussion

Within a multicenter setting, we developed and externally validated an explainable multimodal deep learning framework integrating clinical variables and tumor-centered CT imaging to predict response to neoadjuvant imatinib in patients with GISTs. The bidirectional cross-attention Transformer enabled joint modeling of heterogeneous modalities at the token level, leveraging both intra- and inter-modality interactions to derive patient-level predictions. In internal validation, cross-attention models achieved the highest discrimination, outperforming both clinical-only and imaging-only approaches. However, performance declined in external validation across all models, with convergence of multimodal and unimodal performance. Clinical-only models showed stable performance across settings, whereas imaging-only models were inconsistent. Training from scratch yielded higher internal performance, while self-supervised pretraining resulted in more stable, though not superior, external performance.

Our explainability analyses identified molecular and clinical determinants of response that are broadly consistent with established GIST biology while also providing additional granularity on host-related factors. The higher importance of CD117 and PDGFRA in responders aligns with prior genomic studies demonstrating that activating mutations in these oncogenic drivers are the principal mediators of imatinib sensitivity[28–30]. BRAF also exhibited increased importance in responders in our model, contrasting with reports that BRAF-mutant GISTs are typically KIT/PDGFRA–wild-type and less responsive to imatinib[31], potentially reflecting cohort-specific effects or higher-order interactions captured by the model. In contrast, features associated with non-response were dominated by patient- and disease-level factors, including advanced status at diagnosis and older age, both of which have been linked to poorer outcomes following neoadjuvant imatinib[32]. Notably, comorbidity-related variables such as diabetes, hypertension, hypercholesterolemia, and prior malignancies also showed higher importance in non-responders, supporting emerging evidence that systemic health and host physiology can significantly modulate response to targeted therapies[33].

Our findings support the hypothesis that baseline CT tumor images encode treatment-relevant phenotypic information in GIST, but also indicate that imaging alone is insufficiently robust for reliable response prediction across centers. This is consistent with prior work showing that CT-based assessment of GIST response is informative, particularly when changes in tumor density, morphology, or volume are considered rather than size alone[34]. More recent radiomics and radiogenomic studies further suggest that CT features can capture biological heterogeneity, including mutation patterns linked to imatinib sensitivity, supporting the biological plausibility of an imaging-derived predictive signal[35, 36]. However, the limited external performance of our imaging-only models highlights an important translational limitation, as CT-derived features may be sensitive to center-specific acquisition and reconstruction differences. In contrast, the cross-attention models achieved the strongest internal performance and more balanced external sensitivity and specificity, indicating that CT tumor representations are most informative when interpreted jointly with clinical and molecular variables rather than in isolation. Notably, imaging-only models exhibited pronounced instability, ranging from degenerate behavior in the fine-tuned setting to near-perfect discrimination when trained from scratch in internal validation, highlighting susceptibility to overfitting and potential reliance on shortcut or non-transferable features.

The comparison between self-supervised pretraining followed by LoRA-based fine-tuning and training from scratch highlights a nuanced trade-off between apparent performance and generalizability. Although cross-attention models trained from scratch achieved the highest internal discrimination, this likely reflects reliance on dataset-specific or non-transferable features, a known limitation in medical imaging without pretraining[37]. Such behavior is consistent with shortcut learning, where models exploit spurious signals that do not generalize across datasets[38]. In contrast, fine-tuned models demonstrated more constrained optimization behavior and consistently lower internal AUCs, yet yielded more balanced sensitivity–specificity profiles and comparable discrimination in external validation. This pattern suggests that pretraining acts as an implicit regularizer, stabilizing representation learning and improving transferability[21]. Notably, the benefit of pretraining was most evident in multimodal settings, where fine-tuned cross-attention models avoided the degenerate behavior observed in imaging-only models and maintained more consistent performance across centers. Together, these findings suggest that while training from scratch can maximize in-sample performance, self-supervised pretraining provides more robust representations for generalizable multimodal prediction.

The discrepancy between internal and external performance may be partly explained by both methodological and cohort-related factors. Although internal performance was estimated using out-of-fold predictions, hyperparameter optimization and configuration selection were conducted on the same development cohort using cross-validation AUC as the objective. This introduces the potential for selection-induced bias, whereby top-performing configurations are preferentially retained based on favorable validation fluctuations, leading to optimistic internal estimates despite the absence of direct data leakage. In parallel, differences in clinical and molecular characteristics between the development and external cohorts may have further affected generalizability. Notably, the external cohort exhibited a lower mitotic index, reduced prevalence of KIT mutations, and a higher proportion of missing molecular data, alongside a trend toward differences in histological subtype distribution. Given that these variables are closely linked to GIST biology and imatinib sensitivity, such shifts may alter feature–outcome relationships across cohorts.

Future work may further strengthen the clinical and biological relevance of this framework along two key directions. First, extending the multimodal design to incorporate molecular profiling and digital pathology is a logical next step. Molecular features, including genomic alterations, directly govern imatinib sensitivity and resistance, while whole-slide pathology images provide high-resolution characterization of tumor architecture, cellular morphology, and spatial heterogeneity. Together, these modalities offer complementary, biologically grounded information that is not captured by radiology alone, and their integration with CT-derived features may enable a unified, multi-scale representation of GIST biology. Second, incorporating early on-treatment imaging to capture temporal changes in tumor characteristics may further improve prediction by modeling treatment-induced dynamics. Quantifying changes in density, morphology, or volume through delta-based approaches has shown potential to detect early therapeutic effects and may provide a more sensitive indicator of imatinib response than baseline imaging alone.

Following limitations should be considered when interpreting these findings. First, the retrospective design and modest sample size, particularly in the external test cohort, may limit statistical power and increase susceptibility to sampling variability, despite the multicenter setting. Second, hyperparameter optimization and model selection were conducted on the same development cohort, introducing a risk of selection-induced bias and potentially optimistic internal performance estimates. Third, the observed domain shifts in clinical and molecular characteristics, together with variability in CT acquisition and reconstruction, may have affected model transferability across centers.

In conclusion, this study demonstrates the potential of an explainable cross-attention–based multimodal deep learning framework for predicting response to neoadjuvant imatinib in GIST using routinely available clinical and CT imaging data. Multimodal integration enabled strong internal discrimination and provided biologically meaningful insights through unified modeling of complementary data sources. With further integration of additional biological modalities, this framework has the potential to serve as a clinically relevant tool to support personalized treatment decision-making in GIST.

**Table 1.** Descriptive statistics of the study cohort for prediction of response to neoadjuvant imatinib

| Variable | Total (n=213) | Response (n=116) | Non-response (n=97) | P-value |
|---|---|---|---|---|
| Center | | | | 0.7 |
| Antoni van Leeuwenhoek | 118 (55.4) | 67 (57.8) | 51 (52.6) | |
| Erasmus MC Cancer Institute | 48 (22.5) | 23 (19.8) | 25 (25.8) | |
| Radboud University Medical Center | 26 (12.2) | 15 (12.9) | 11 (11.3) | |
| Leiden University Medical Center | 21 (9.9) | 11 (9.5) | 10 (10.3) | |
| Age at treatment initiation | 65 [56, 73] | 63.5 [56, 71] | 66 [57, 74] | 0.086 |
| Sex | | | | 0.081 |
| Female | 94 (44.1) | 58 (50) | 36 (37.1) | |
| Male | 119 (55.9) | 58 (50) | 61 (62.9) | |
| Primary tumor site | | | | 0.2 |
| Gastric | 150 (70.4) | 85 (73.3) | 65 (67) | |
| Small bowel | 24 (11.3) | 14 (12.1) | 10 (10.3) | |
| Duodenal | 23 (10.8) | 10 (8.6) | 13 (13.4) | |
| Esophagus | 9 (4.2) | 6 (5.2) | 3 (3.1) | |
| Other | 7 (3.2) | 1 (0.9) | 6 (6.2) | |
| Status at diagnosis | | | | 0.003 |
| Localized disease | 21 (9.9) | 5 (4.3) | 16 (16.5) | |
| Locally advanced | 187 (87.8) | 110 (94.8) | 77 (79.4) | |
| Metastasized | 5 (2.3) | 1 (0.9) | 4 (4.1) | |
| Longest axial diameter (mm) | 100 [70,148] | 112 [76,160] | 89 [60,140] | 0.026 |
| Mitotic index | 1 [0,5] | 3 [0,9.2] | 0 [0,1] | <0.001 |
| Missing | 90 (42.2) | 44 (37.9) | 46 (47.4) | |
| Histology | | | | 0.3 |
| Spindle cell | 157 (73.7) | 89 (76.7) | 68 (70.1) | |
| Epithelioid | 11 (5.2) | 5 (4.3) | 6 (6.2) | |
| Mixed type | 15 (7) | 5 (4.3) | 10 (10.3) | |
| Missing | 30 (14.1) | 17 (14.7) | 13 (13.4) | |
| CD117 | | | | 0.9 |
| Positive | 174 (81.7) | 94 (81) | 80 (82.5) | |
| Negative | 4 (1.9) | 2 (1.7) | 2 (2.1) | |
| Missing | 35 (16.4) | 20 (17.2) | 15 (15.5) | |
| DOG1 | | | | 0.9 |
| Positive | 141 (66.2) | 76 (65.5) | 65 (67) | |
| Missing | 72 (33.8) | 40 (34.5) | 32 (33) | |
| KIT mutation | | | | 0.019 |
| Present | 135 (63.4) | 80 (69) | 55 (56.7) | |
| Absent | 27 (12.7) | 8 (6.9) | 19 (19.6) | |
| Missing | 51 (23.9) | 28 (24.1) | 23 (23.7) | |
| PDGFRA mutation | | | | 0.1 |
| Present | 19 (8.9) | 6 (5.2) | 13 (13.4) | |
| Absent | 114 (53.5) | 65 (56) | 49 (50.5) | |
| Missing | 80 (37.6) | 45 (38.8) | 35 (36.1) | |
| BRAF mutation | | | | 0.7 |

| Absent | 97 (45.5) | 51 (44) | 46 (47.4) | |
|---|---|---|---|---|
| Missing | 116 (54.5) | 65 (56) | 51 (52.6) | |
| Diabetes | 21 (9.9) | 14 (12.1) | 7 (7.2) | 0.3 |
| Hypertension | 46 (21.6) | 29 (25) | 17 (17.5) | 0.2 |
| Hypercholesterolemia | 15 (7) | 8 (6.9) | 7 (7.2) | 0.9 |
| Other malignancies | 60 (28.2) | 29 (25) | 31 (32) | 0.3 |

**Notes:** Values presented as median [Q1,Q3] or n (%). P-values were calculated using the Mann–Whitney U test, Chi-square test, or Fisher's exact test.

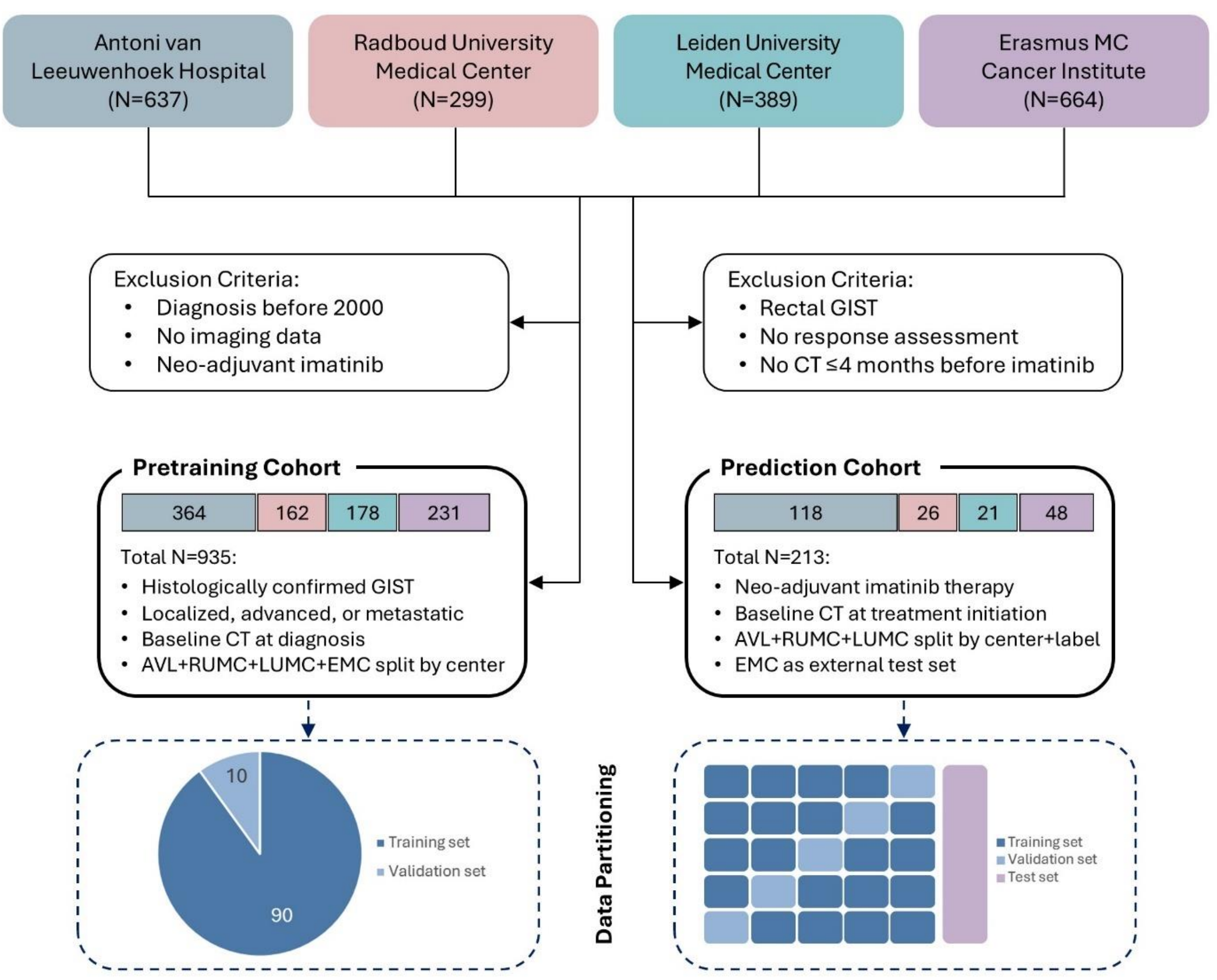

**Figure 1.** Schematic of multicenter patient selection, cohort derivation, and data partitioning for pretraining and response prediction to neoadjuvant imatinib in gastrointestinal stromal tumors.

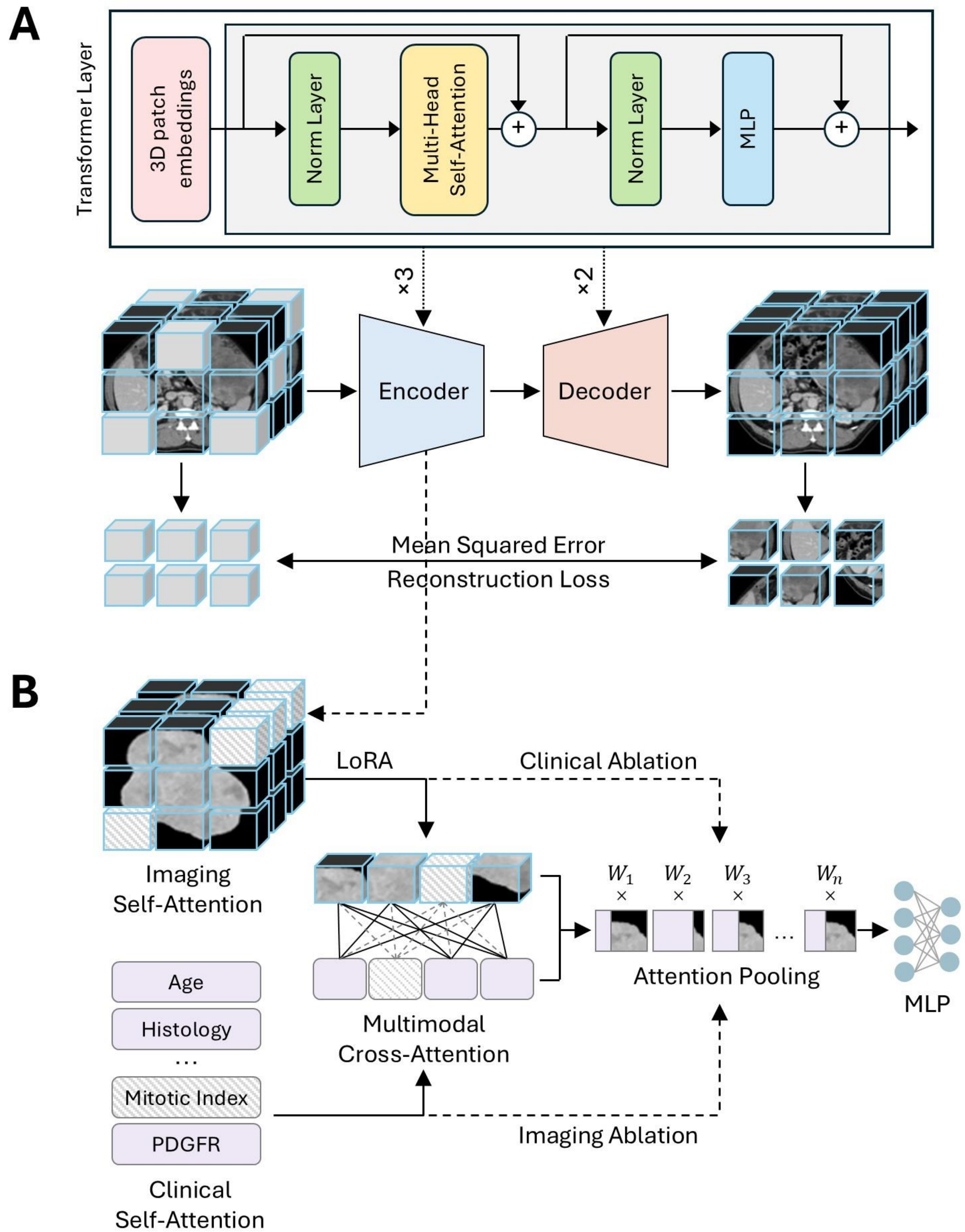

**Figure 2.** Overview of analysis framework. **A)** Masked autoencoder pretraining of the imaging encoder using abdominal computed tomography in patients with gastrointestinal stromal tumors. **B)** Multimodal transformer-based cross-attention framework for predicting response to neoadjuvant imatinib through integration of clinical and CT-derived imaging information, with explicit missing-value and stochastic masking of clinical tokens, padding-aware masking of imaging tokens, and unimodal ablation architectures excluding cross-attention and complementary modality integration. **Abbreviations:** LoRA, Low-Rank Adaptation; MLP, Multilayer Perceptron.

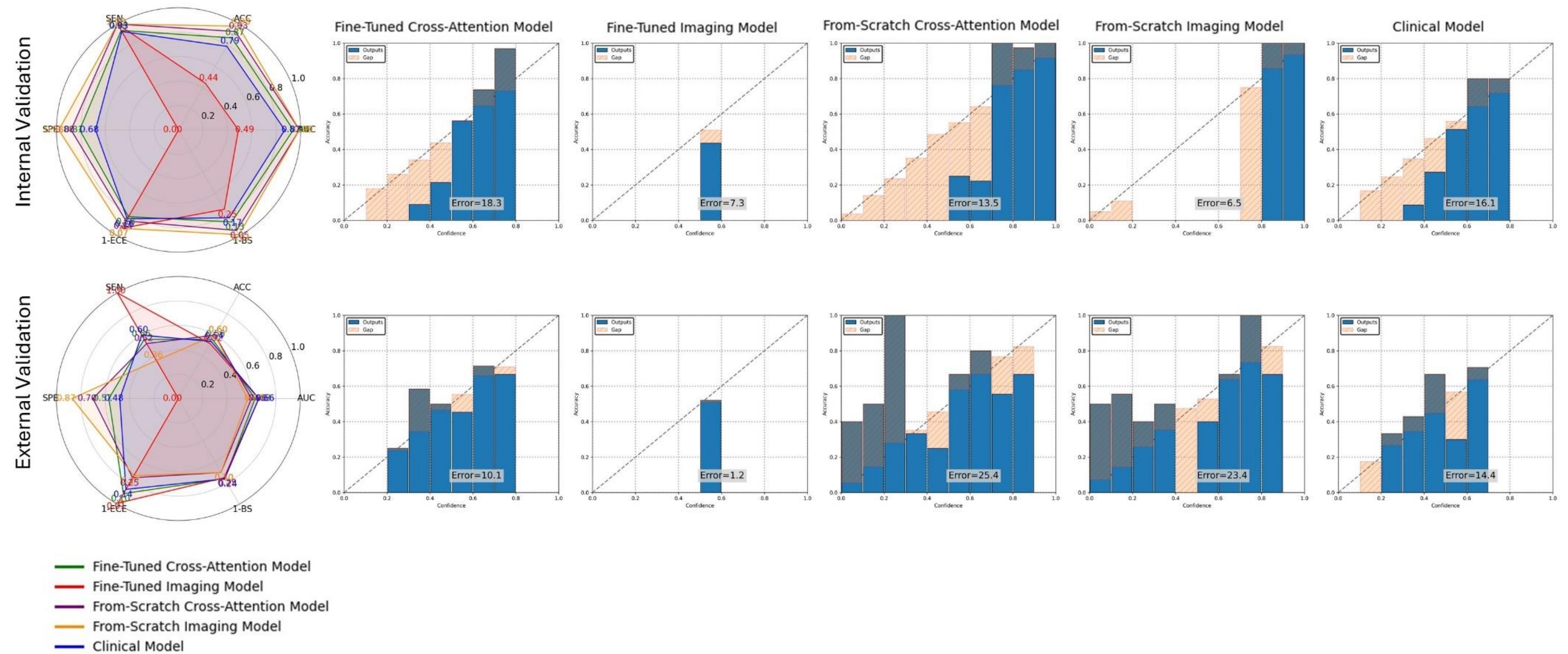

**Figure 3.** Discrimination (area under the receiver operating characteristic curve, accuracy, sensitivity, and specificity) and calibration (expected calibration error, Brier score, and reliability diagrams) of multimodal and unimodal models in internal and external validation for predicting response to neoadjuvant imatinib in gastrointestinal stromal tumors.

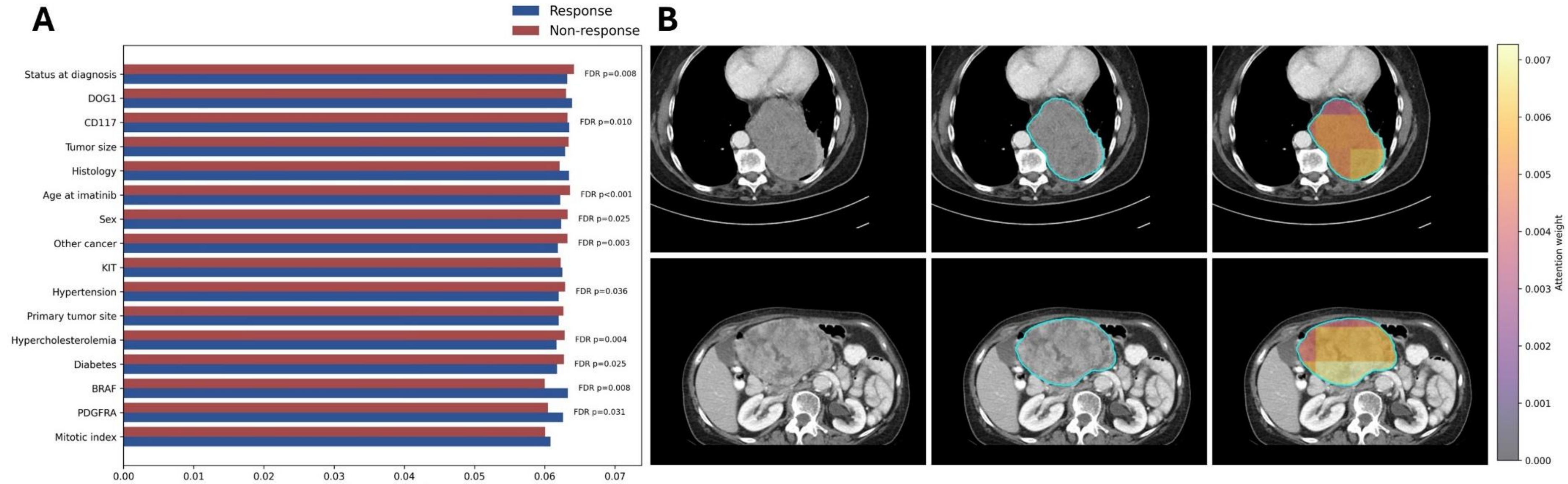

**Figure 4.** Explainability of multimodal predictions. **A)** Clinical feature importance analysis in the development cohort, comparing responders and non-responders, with significance assessed using Mann–Whitney U tests and false discovery rate correction. **B)** Representative external true negative (non-response, top) and true positive (response, bottom) cases with CT, tumor segmentation, and attention overlays highlighting regions driving predictions.

**Table S1.** Hyperparameter search space explored during SMAC optimization for the fine-tuned cross-attention multimodal model with LoRA

| No. | Hyperparameter | Search range |
|---|---|---|
| 1 | Initial learning rate | Log-uniform range: $5 \times 10^{-5}$ to $5 \times 10^{-4}$ |
| 2 | AdamW weight decay | Log-uniform range: $1 \times 10^{-5}$ to $3 \times 10^{-4}$ |
| 3 | Clinical self-attention encoder depth | 1, 2 layers |
| 4 | Number of attention heads in the clinical encoder | 2, 4 |
| 5 | Cross-attention fusion depth | 1, 2 layers |
| 6 | Number of attention heads in the cross-attention module | 1, 2, 4 |
| 7 | Classifier head architecture | Single-layer MLP; two-layer MLP |
| 8 | Minimum tumor-fraction threshold for retaining imaging patches | 0.20, 0.30, 0.40 |
| 9 | Clinical feature dropout probability | 0.00, 0.10 |
| 10 | Random rotation augmentation range for CT imaging | 0°, ±5°, ±10° |
| 11 | LoRA rank | 2, 4, 8 |
| 12 | LoRA scaling ratio | 0.5, 1.0, 2.0 |

**Table S2.** Hyperparameter search space explored during SMAC optimization for the fine-tuned imaging-only model with LoRA

| No. | Hyperparameter | Search range |
|---|---|---|
| 1 | Initial learning rate | Log-uniform range: $5 \times 10^{-5}$ to $5 \times 10^{-4}$ |
| 2 | AdamW weight decay | Log-uniform range: $1 \times 10^{-5}$ to $3 \times 10^{-4}$ |
| 3 | Classifier head architecture | Single-layer MLP; two-layer MLP |
| 4 | Minimum tumor-fraction threshold for retaining imaging patches | 0.20, 0.30, 0.40 |
| 5 | Random rotation augmentation range for CT imaging | 0°, ±5°, ±10° |
| 6 | LoRA rank | 2, 4, 8 |
| 7 | LoRA scaling ratio | 0.5, 1.0, 2.0 |

**Table S3.** Hyperparameter search space explored during SMAC optimization for the clinical-only model

| No. | Hyperparameter | Search range |
|---|---|---|
| 1 | Initial learning rate | Log-uniform range: $5 \times 10^{-5}$ to $5 \times 10^{-4}$ |
| 2 | AdamW weight decay | Log-uniform range: $1 \times 10^{-5}$ to $3 \times 10^{-4}$ |
| 3 | Clinical self-attention encoder depth | 1, 2, 3 layers |
| 4 | Number of attention heads in the clinical encoder | 2, 4 |
| 5 | Classifier head architecture | Single-layer MLP; two-layer MLP |
| 6 | Clinical feature dropout probability | 0.00, 0.10 |

**Table S4.** Hyperparameter search space explored during SMAC optimization for the cross-attention multimodal model trained from scratch

| No. | Hyperparameter | Search range |
|---|---|---|
| 1 | Imaging patch size | 12 × 48 × 48; 8 × 64 × 64; 16 × 32 × 32; 16 × 64 × 64 voxels |
| 2 | Token embedding dimension | 64, 128, 256 |
| 3 | Initial learning rate | Log-uniform range: $3 \times 10^{-5}$ to $5 \times 10^{-4}$ |
| 4 | AdamW weight decay | Log-uniform range: $1 \times 10^{-6}$ to $5 \times 10^{-3}$ |

| No. | Hyperparameter | Search range |
|---|---|---|
| 5 | Clinical self-attention encoder depth | 1, 2 layers |
| 6 | Number of attention heads in the clinical encoder | 2, 4 |
| 7 | Imaging self-attention encoder depth | 2, 3 layers |
| 8 | Number of attention heads in the imaging encoder | 4, 8 |
| 9 | Cross-attention fusion depth | 1, 2 layers |
| 10 | Number of attention heads in the cross-attention module | 2, 4 |
| 11 | Classifier head architecture | Single-layer MLP; two-layer MLP |
| 12 | Minimum tumor-fraction threshold for retaining imaging patches | 0.20, 0.30, 0.40 |
| 13 | Clinical feature dropout probability | 0.00, 0.10 |
| 14 | Random rotation augmentation range for CT imaging | 0°, ±5°, ±10° |

**Table S5.** Hyperparameter search space explored during SMAC optimization for the imaging-only model trained from scratch

| No. | Hyperparameter | Search range |
|---|---|---|
| 1 | Imaging patch size | 12 × 48 × 48; 8 × 64 × 64; 16 × 32 × 32; 16 × 64 × 64 voxels |
| 2 | Token embedding dimension | 64, 128, 256 |
| 3 | Initial learning rate | Log-uniform range: $3 \times 10^{-5}$ to $5 \times 10^{-4}$ |
| 4 | AdamW weight decay | Log-uniform range: $1 \times 10^{-6}$ to $5 \times 10^{-3}$ |
| 5 | Imaging self-attention encoder depth | 2, 3 layers |
| 6 | Number of attention heads in the imaging encoder | 4, 8 |
| 7 | Classifier head architecture | Single-layer MLP; two-layer MLP |
| 8 | Minimum tumor-fraction threshold for retaining imaging patches | 0.20, 0.30, 0.40 |
| 9 | Random rotation augmentation range for CT imaging | 0°, ±5°, ±10° |

**Table S6.** Descriptive statistics of the development (AVL, RUMC, LUMC) and external test (EMC) datasets

| Variable | Development (n=165) | External test (n=48) | P-value |
|---|---|---|---|
| Treatment response | | | 0.4 |
| Non-response | 93 (56.4) | 23 (47.9) | |
| Response | 72 (43.6) | 25 (52.1) | |
| Age at treatment initiation | 64 [56,73] | 65 [59,72] | 0.6 |
| Sex | | | 0.2 |
| Female | 68 (41.2) | 26 (54.2) | |
| Male | 97 (58.8) | 22 (45.8) | |
| Primary tumor site | | | 0.3 |
| Gastric | 117 (70.9) | 33 (68.8) | |
| Small bowel | 20 (12.1) | 4 (8.3) | |
| Duodenal | 14 (8.5) | 9 (18.8) | |
| Esophagus | 7 (4.2) | 2 (4.2) | |
| Other | 7 (4.2) | 0 (0) | |
| Status at diagnosis | | | 0.9 |
| Localized disease | 16 (9.7) | 5 (10.4) | |
| Locally advanced | 145 (87.9) | 42 (87.5) | |
| Metastasized | 4 (2.4) | 1 (2.1) | |
| Longest axial diameter (mm) | 102.5 [70,152.2] | 86 [67.5,130] | 0.2 |
| Mitotic index | 1 [0,6] | 0 [0,1.8] | 0.043 |
| Missing | 64 (38.8) | 26 (54.2) | |
| Histology | | | 0.087 |
| Spindle cell | 127 (77) | 30 (62.5) | |
| Epithelioid | 9 (5.5) | 2 (4.2) | |
| Mixed type | 11 (6.7) | 4 (8.3) | |
| Missing | 18 (10.9) | 12 (25) | |
| CD117 | | | 0.5 |
| Positive | 135 (81.8) | 39 (81.2) | |
| Negative | 4 (2.4) | 0 (0) | |
| Missing | 26 (15.8) | 9 (18.8) | |
| DOG1 | | | 0.3 |
| Positive | 113 (68.5) | 28 (58.3) | |
| Missing | 52 (31.5) | 20 (41.7) | |
| KIT mutation | | | 0.044 |
| Present | 110 (66.7) | 25 (52.1) | |
| Absent | 22 (13.3) | 5 (10.4) | |
| Missing | 33 (20) | 18 (37.5) | |
| PDGFRA mutation | | | 0.2 |
| Present | 18 (10.9) | 1 (2.1) | |
| Absent | 87 (52.7) | 27 (56.2) | |
| Missing | 60 (36.4) | 20 (41.7) | |
| BRAF mutation | | | 0.7 |
| Absent | 77 (46.7) | 20 (41.7) | |
| Missing | 88 (53.3) | 28 (58.3) | |
| Diabetes | 17 (10.3) | 4 (8.3) | 0.8 |
| Hypertension | 37 (22.4) | 9 (18.8) | 0.7 |
| Hypercholesterolemia | 12 (7.3) | 3 (6.2) | 0.9 |
| Other malignancies | 46 (27.9) | 14 (29.2) | 0.9 |

**Notes:** Values presented as median [Q1,Q3] or n (%). P-values were calculated using the Mann–Whitney U test, Chi-square test, or Fisher's exact test.

**Table S7.** Top five hyperparameter configurations selected for ensembling in the cross-attention model with LoRA fine-tuning

| Config ID | Mean Val AUC | Learning Rate | Weight Decay | Clinical Depth | Clinical Heads | Cross-Attn Depth | Cross-Attn Heads | Classifier Head | Min Tumor Fraction | Clinical Dropout | Imaging Rotation | LoRA Rank (r) | LoRA Ratio |
|---|---|---|---|---|---|---|---|---|---|---|---|---|---|
| 57 | 0.750 | $1.93\times10^{-4}$ | $6.67\times10^{-5}$ | 2 | 2 | 1 | 4 | Two-layer | 0.20 | 0.10 | ±10° | 4 | 1.0 |
| 26 | 0.747 | $3.39\times10^{-4}$ | $3.47\times10^{-5}$ | 2 | 2 | 1 | 4 | Two-layer | 0.20 | 0.10 | ±10° | 4 | 0.5 |
| 38 | 0.744 | $1.25\times10^{-4}$ | $4.87\times10^{-5}$ | 2 | 2 | 2 | 1 | Two-layer | 0.20 | 0.10 | ±10° | 4 | 0.5 |
| 35 | 0.744 | $3.94\times10^{-4}$ | $7.09\times10^{-5}$ | 2 | 2 | 1 | 4 | Two-layer | 0.20 | 0.10 | ±10° | 8 | 1.0 |
| 33 | 0.737 | $1.66\times10^{-4}$ | $4.44\times10^{-5}$ | 2 | 2 | 1 | 2 | Single-layer | 0.20 | 0.00 | ±5° | 2 | 1.0 |

**Table S8.** Top five hyperparameter configurations selected for ensembling in the imaging-only model with LoRA fine-tuning

| Config ID | Mean Val AUC | Learning Rate | Weight Decay | Classifier Head | Min Tumor Fraction | Imaging Rotation | LoRA Rank (r) | LoRA Ratio |
|---|---|---|---|---|---|---|---|---|
| 37 | 0.525 | $4.56 \times 10^{-4}$ | $2.08 \times 10^{-4}$ | Two-layer | 0.30 | 0° | 8 | 1.0 |
| 50 | 0.518 | $4.46 \times 10^{-4}$ | $2.87 \times 10^{-4}$ | Two-layer | 0.20 | 0° | 4 | 2.0 |
| 22 | 0.512 | $6.59 \times 10^{-5}$ | $1.26 \times 10^{-5}$ | Two-layer | 0.20 | ±5° | 2 | 1.0 |
| 42 | 0.511 | $6.58 \times 10^{-5}$ | $1.94 \times 10^{-4}$ | Two-layer | 0.20 | ±10° | 4 | 1.0 |
| 46 | 0.510 | $2.14 \times 10^{-4}$ | $2.20 \times 10^{-4}$ | Two-layer | 0.20 | ±10° | 2 | 0.5 |

**Table S9.** Top five hyperparameter configurations selected for ensembling in the clinical-only model

| Config ID | Mean Val AUC | Learning Rate | Weight Decay | Clinical Encoder Depth | Attention Heads | Classifier Head | Clinical Augmentation |
|---|---|---|---|---|---|---|---|
| 53 | 0.696 | $4.23 \times 10^{-4}$ | $1.62 \times 10^{-4}$ | 3 | 2 | Two-layer | 0.10 |
| 13 | 0.689 | $3.62 \times 10^{-4}$ | $1.81 \times 10^{-4}$ | 2 | 4 | Single-layer | 0.00 |
| 15 | 0.686 | $4.56 \times 10^{-4}$ | $1.08 \times 10^{-4}$ | 3 | 4 | Two-layer | 0.10 |
| 23 | 0.686 | $4.21 \times 10^{-4}$ | $9.00 \times 10^{-5}$ | 3 | 2 | Single-layer | 0.10 |
| 36 | 0.680 | $2.25 \times 10^{-4}$ | $1.95 \times 10^{-4}$ | 3 | 4 | Single-layer | 0.10 |

**Table S10.** Top five hyperparameter configurations selected for ensembling in the cross-attention model trained from scratch

| Config ID | Mean Val AUC | Patch Size | Embedding dimension | Learning Rate | Weight Decay | Clinical Depth | Clinical Heads | Imaging Depth | Imaging Heads | Cross-Attn Depth | Cross-Attn Heads | Classifier Head | Min Tumor Fraction | Clinical Dropout | Imaging Rotation |
|---|---|---|---|---|---|---|---|---|---|---|---|---|---|---|---|
| 57 | 0.788 | 8×64×64 | 256 | $5.00 \times 10^{-4}$ | $1.44 \times 10^{-3}$ | 2 | 4 | 3 | 4 | 2 | 4 | Single-layer | 0.30 | 0.00 | ±5° |
| 32 | 0.753 | 8×64×64 | 256 | $5.00 \times 10^{-4}$ | $9.56 \times 10^{-6}$ | 2 | 4 | 3 | 4 | 2 | 4 | Two-layer | 0.30 | 0.10 | 0° |
| 54 | 0.748 | 8×64×64 | 256 | $5.00 \times 10^{-4}$ | $5.98 \times 10^{-5}$ | 2 | 4 | 2 | 8 | 1 | 2 | Two-layer | 0.30 | 0.00 | ±5° |
| 33 | 0.736 | 16×32×32 | 256 | $4.96 \times 10^{-4}$ | $9.58 \times 10^{-6}$ | 2 | 4 | 3 | 8 | 2 | 2 | Single-layer | 0.30 | 0.10 | 0° |
| 53 | 0.734 | 8×64×64 | 256 | $5.00 \times 10^{-4}$ | $8.72 \times 10^{-6}$ | 2 | 4 | 2 | 8 | 1 | 2 | Single-layer | 0.30 | 0.10 | ±5° |

**Table S11.** Top five hyperparameter configurations selected for ensembling in the imaging-only model trained from scratch

| Config ID | Mean Val AUC | Patch Size | Embedding dimension | Learning Rate | Weight Decay | Imaging Depth | Imaging Heads | Classifier Head | Min Tumor Fraction | Imaging Rotation |
|---|---|---|---|---|---|---|---|---|---|---|
| 42 | 0.693 | 16×32×32 | 128 | $2.09 \times 10^{-4}$ | $2.84 \times 10^{-3}$ | 2 | 8 | Single-layer | 0.30 | ±5° |
| 58 | 0.667 | 16×32×32 | 128 | $2.23 \times 10^{-4}$ | $2.95 \times 10^{-3}$ | 3 | 8 | Single-layer | 0.30 | ±10° |
| 56 | 0.656 | 16×32×32 | 128 | $2.14 \times 10^{-4}$ | $2.92 \times 10^{-3}$ | 2 | 8 | Single-layer | 0.30 | ±10° |
| 50 | 0.649 | 16×32×32 | 128 | $2.08 \times 10^{-4}$ | $3.21 \times 10^{-3}$ | 2 | 8 | Single-layer | 0.30 | ±5° |
| 46 | 0.634 | 16×32×32 | 128 | $2.09 \times 10^{-4}$ | $2.84 \times 10^{-3}$ | 2 | 8 | Single-layer | 0.30 | 0° |

**Table S12.** Internal and external validation performance of fine-tuned and from-scratch multimodal and unimodal models for predicting response to neoadjuvant imatinib in gastrointestinal stromal tumors

| Internal Validation | AUC | Accuracy | Sensitivity | Specificity | Brier Score | ECE |
|---|---|---|---|---|---|---|
| Fine-Tuned Cross-Attention Model | 0.94 (0.91–0.97) | 0.87 (0.81–0.92) | 0.94 (0.89–0.99) | 0.81 (0.72–0.89) | 0.13 (0.12–0.15) | 0.18 (0.15–0.24) |
| Fine-Tuned Imaging Model | 0.49 (0.40–0.57) | 0.44 (0.44–0.44) | 1.00 (1.00–1.00) | 0.00 (0.00–0.00) | 0.25 (0.25–0.25) | 0.07 (0.07–0.07) |
| From-Scratch Cross-Attention Model | 0.99 (0.99–1.00) | 0.93 (0.89–0.97) | 1.00 (1.00–1.00) | 0.88 (0.81–0.95) | 0.05 (0.04–0.07) | 0.14 (0.11–0.16) |
| From-Scratch Imaging Model | 1.00 (1.00–1.00) | 0.99 (0.98–1.00) | 1.00 (1.00–1.00) | 0.99 (0.97–1.00) | 0.01 (0.00–0.02) | 0.07 (0.06–0.08) |
| Clinical Model | 0.87 (0.82–0.92) | 0.79 (0.73–0.85) | 0.93 (0.86–0.99) | 0.68 (0.58–0.77) | 0.17 (0.15–0.19) | 0.16 (0.12–0.22) |
| | | | | | | |
| **External Validation** | **AUC** | **Accuracy** | **Sensitivity** | **Specificity** | **Brier Score** | **ECE** |
| Fine-Tuned Cross-Attention Model | 0.63 (0.47–0.78) | 0.56 (0.42–0.69) | 0.56 (0.36–0.76) | 0.57 (0.35–0.78) | 0.24 (0.20–0.28) | 0.10 (0.07–0.27) |
| Fine-Tuned Imaging Model | 0.66 (0.50–0.81) | 0.52 (0.52–0.52) | 1.00 (1.00–1.00) | 0.00 (0.00–0.00) | 0.25 (0.25–0.25) | 0.01 (0.01–0.01) |
| From-Scratch Cross-Attention Model | 0.60 (0.43–0.75) | 0.60 (0.46–0.73) | 0.52 (0.32–0.72) | 0.70 (0.52–0.87) | 0.30 (0.22–0.39) | 0.25 (0.18–0.41) |
| From-Scratch Imaging Model | 0.56 (0.40–0.72) | 0.56 (0.44–0.69) | 0.40 (0.20–0.60) | 0.74 (0.57–0.91) | 0.30 (0.23–0.37) | 0.30 (0.23–0.37) |
| Clinical Model | 0.66 (0.50–0.80) | 0.54 (0.40–0.69) | 0.60 (0.40–0.80) | 0.48 (0.26–0.70) | 0.24 (0.20–0.27) | 0.14 (0.08–0.28) |

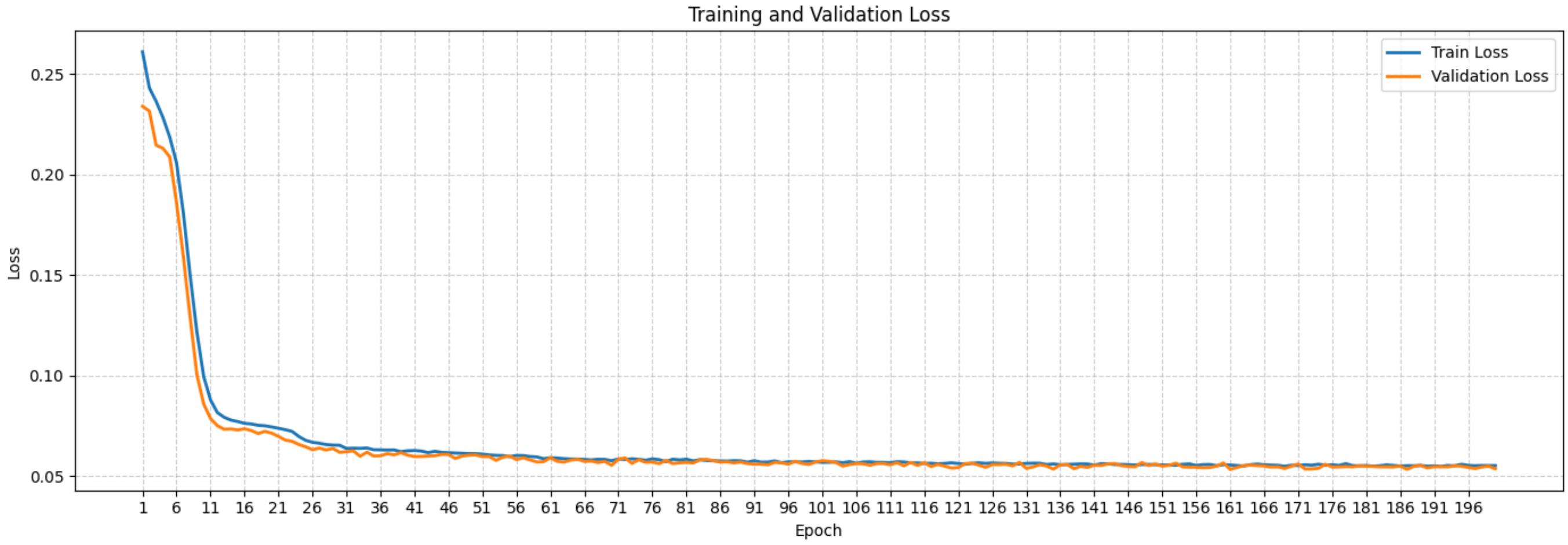

**Figure S1.** Training and validation AUCs during masked autoencoder (MAE) pretraining of the imaging encoder, using a center-stratified 90/10 split, illustrating convergence and generalization across epochs.

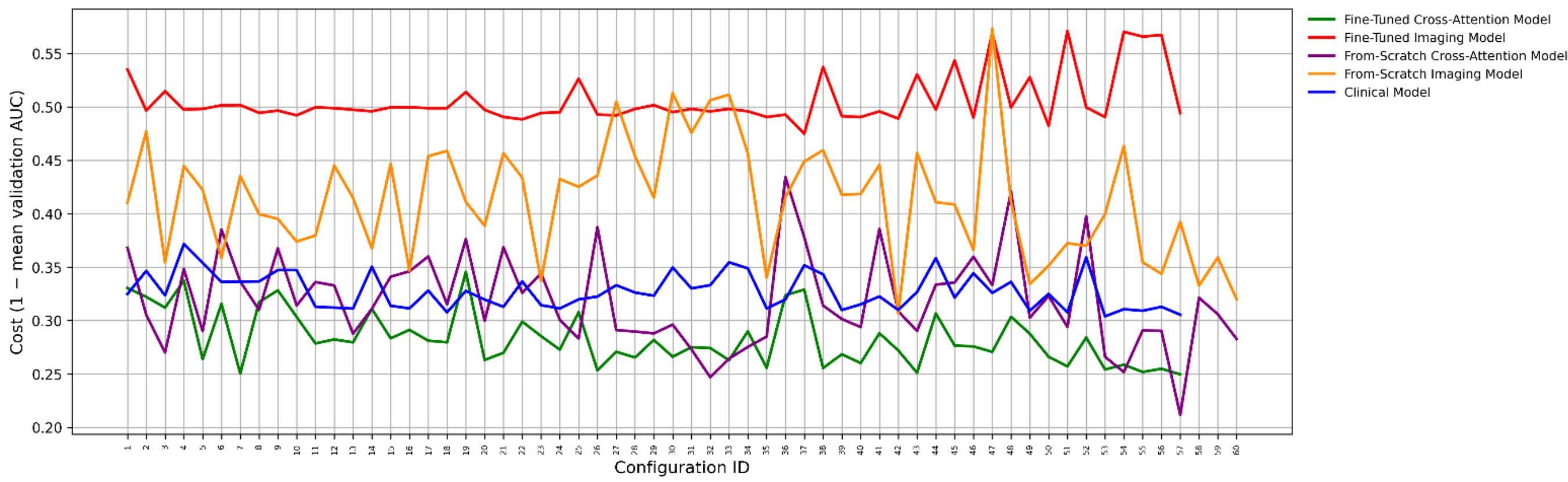

**Figure S2.** SMAC3 hyperparameter optimization cost values for the fine-tuned cross-attention, fine-tuned imaging, from-scratch cross-attention, from-scratch imaging, and clinical models, computed from the mean validation AUC across 5-fold cross-validation on the development cohort.